\documentclass[prl,reprint,shownopacs,superscriptaddress]{revtex4-2}

\usepackage{graphicx}
\usepackage{amssymb,amsmath,amsthm,bm}
\usepackage{lipsum}
\usepackage[colorlinks=true,linkcolor=blue,citecolor=blue,pdfauthor={ },pdftitle={ },pdfsubject={ },pdfkeywords={ }]{hyperref}
\usepackage{xcolor}
\graphicspath{{Figures/}}
\usepackage{times}

\usepackage[small,raggedright]{titlesec}
\titleformat*{\section} {\small\bf}
\titlespacing*{\section} {0pt}{10pt}{0pt}
\titlespacing*{\subsection} {0pt}{10pt}{0pt}

\usepackage{graphicx}
\usepackage{color}
\usepackage{lineno}
\definecolor{dred}{rgb}{.8,0.2,.2}
\definecolor{ddred}{rgb}{.8,0.5,.5}
\definecolor{dblue}{rgb}{.2,0.2,.8}
\definecolor{dgreen}{rgb}{.2,0.5,.2}

\newcommand{\bra}[1]{\mbox{$\langle #1|$}}
\newcommand{\ket}[1]{\ensuremath{|#1\rangle}}

\begin{document}

	\title{Noise-Resilient Quantum Metrology with Quantum Computing}
	
	\author{Xiangyu Wang}
	\thanks{These authors contributed equally to this work}
	\affiliation{Department of Physics, State Key Laboratory of Quantum Functional Materials, and Guangdong Basic Research Center of Excellence for Quantum Science, Southern University of Science and Technology, Shenzhen 518055, China}
    \affiliation{College of Metrology Measurement and Instrument, China Jiliang University, Hangzhou, 310018, China}
	
	\author{Chenrong Liu}
	\thanks{These authors contributed equally to this work}
	\affiliation{College of Metrology Measurement and Instrument, China Jiliang University, Hangzhou, 310018, China}

    \author{Xue Lin}
	\affiliation{Anhui  Institute  of  Medicine, Hefei, 230601, China}

    \author{Yu Tian}
	\affiliation{College of Metrology Measurement and Instrument, China Jiliang University, Hangzhou, 310018, China}

    \author{Yishan Li}
	\affiliation{International Quantum Academy, Shenzhen, 518048, China}

    \author{Xinfang Nie}
    \affiliation{Quantum Science Center of Guangdong-HongKong-Macao Greater Bay Area, Shenzhen 518045, China}
    \affiliation{Department of Physics, State Key Laboratory of Quantum Functional Materials, and Guangdong Basic Research Center of Excellence for Quantum Science, Southern University of Science and Technology, Shenzhen 518055, China}

    \author{Yufang Feng}
	\affiliation{Department of Physics, State Key Laboratory of Quantum Functional Materials, and Guangdong Basic Research Center of Excellence for Quantum Science, Southern University of Science and Technology, Shenzhen 518055, China}

    \author{Yuxuan Zheng}
	\affiliation{Department of Physics, State Key Laboratory of Quantum Functional Materials, and Guangdong Basic Research Center of Excellence for Quantum Science, Southern University of Science and Technology, Shenzhen 518055, China}

	\author{Ying Dong}
	\email{yingdong@cjlu.edu.cn}
	\affiliation{College of Metrology Measurement and Instrument, China Jiliang University, Hangzhou, 310018, China}
    
	\author{Xinqing Wang}
	\email{wxqnano@cjlu.edu.cn}
	\affiliation{College of Metrology Measurement and Instrument, China Jiliang University, Hangzhou, 310018, China}
	
	\author{Dawei Lu}
	\email{ludw@sustech.edu.cn}
	\affiliation{Department of Physics, State Key Laboratory of Quantum Functional Materials, and Guangdong Basic Research Center of Excellence for Quantum Science, Southern University of Science and Technology, Shenzhen 518055, China}
	\affiliation{Quantum Science Center of Guangdong-HongKong-Macao Greater Bay Area, Shenzhen 518045, China}

	\date{\today}
	
	\begin{abstract}
		Quantum computing has made remarkable strides in recent years, as demonstrated by quantum supremacy experiments and the realization of high-fidelity, fault-tolerant gates. However, a major obstacle persists: practical real-world applications remain scarce, largely due to the inefficiency of loading classical data into quantum processors. Here, we propose an alternative strategy that shifts the focus from classical data encoding to directly processing quantum data. We target quantum metrology, a practical quantum technology whose precision is often constrained by realistic noise. We develop an experimentally feasible scheme in which a quantum computer optimizes information acquired from quantum metrology, thereby enhancing performance in noisy quantum metrology tasks and overcoming the classical-data-loading bottleneck. We demonstrate this approach through experimental implementation with nitrogen-vacancy centers in diamond and numerical simulations using models of distributed superconducting quantum processors. Our results show that this method improves the accuracy of sensing estimates and significantly boosts sensitivity, as quantified by the quantum Fisher information, thus offering a new pathway to harness near-term quantum computers for realistic quantum metrology.
	\end{abstract}
	
	\maketitle
Quantum computing (QC) has the potential to solve certain problems exponentially faster than classical computers, offering a promising path for next-generation information processing. Recent milestones, such as demonstrations of quantum supremacy \cite{arute2019quantum, zhong2021phase, deng2023gaussian}, error correction \cite{ofek2016extending,egan2021fault,sivak2023real,ni2023beating}, and logical qubit operations \cite{bluvstein2024logical,abobeih2022fault,erhard2021entangling}, are bringing fully fault-tolerant quantum machines closer to reality. However, applying quantum computers to real-world tasks still faces major hurdles, especially the challenge of loading large volumes of classical data into a quantum processor efficiently. Many well-known quantum algorithms \cite{grover1997quantum,harrow2009quantum} assume that the input data is already encoded in a quantum state, but preparing such states using classical resources often incurs exponential costs. Although quantum random access memory has been suggested as a solution \cite{giovannetti2008quantum, park2019circuit}, its practical implementation remains challenging.

Quantum metrology (QM), a key application of quantum technologies, aims to exploit quantum effects to measure weak signals and subtle physical quantities more precisely than classical approaches \cite{giovannetti2011advances, toth2014quantum, pezze2018quantum, barry2020sensitivity}. However, noise from the environment often reduces measurement \textit{accuracy} (how close a result is to the true value) and \textit{precision} (the consistency of repeated measurements). Highly entangled probe states, which are essential for surpassing the standard quantum limit and approaching the Heisenberg limit \cite{leibfried2004toward, mitchell2004super, walther2004broglie, monz201114}, are particularly vulnerable to decoherence in realistic settings \cite{ haase2016precision, aolita2008scaling, joo2011quantum, grun2022protocol}. This makes it challenging to reliably extract useful information from noisy quantum states.

To tackle these challenges, we propose a combined QM+QC strategy that integrates QM and QC to boost both data handling and noise resilience. In this approach, the output from a quantum sensor, which carries quantum information in the form of a mixed state, is not directly measured as in conventional schemes. Instead, it is transferred to a more stable quantum processor. The quantum processor then applies quantum machine learning techniques to refine and analyze the noisy data, enabling step-by-step improvement in measurement accuracy and precision.


We validate our method with both experiments and simulations. In our demonstration with nitrogen-vacancy (NV) centers in diamond, we measure a magnetic field while deliberately adding varying levels of noise, and find that QM+QC enhances the measurement accuracy by 200 times even under strong noise conditions. We also simulate a two-module distributed superconducting quantum system for sensing a magnetic field under noise, where each module has four qubits: one module as the sensor and the other as the processor. The results show that after applying QM+QC, the quantum Fisher information (QFI), which indicates precision, improves by 13.27 dB and gets much closer to the Heisenberg limit.
These results show that combining QM with QC holds promise for practical, noise-resilient sensing and could broaden the use of quantum computers in real-world applications.
	
	\begin{figure*}
		\centering
		\includegraphics[width=0.9\textwidth]{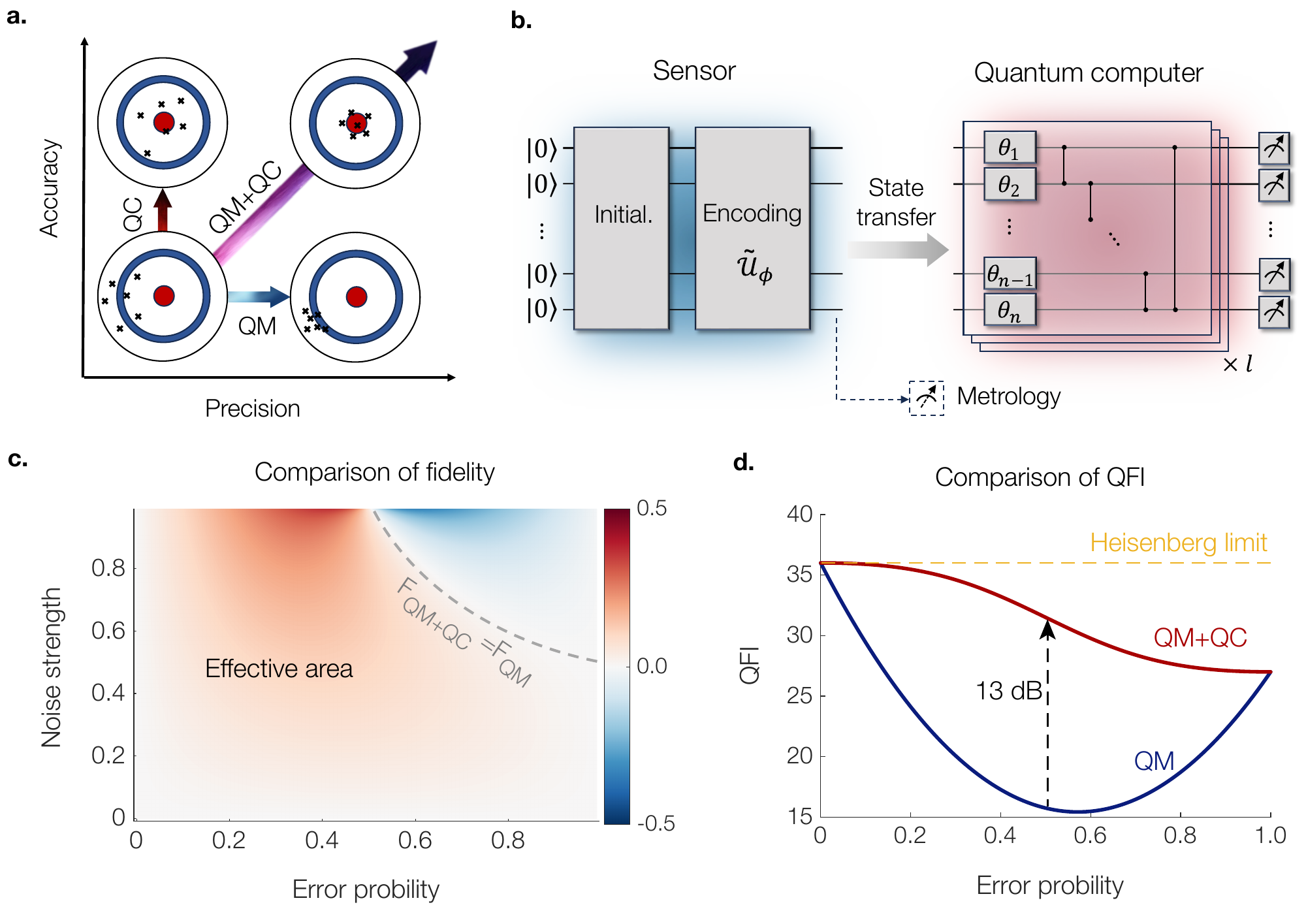}
		\caption{\textbf{Concept and theoretical framework of the QM+QC method.} \textbf{a,} Conceptual illustration of accuracy and precision in statistics, represented by dartboard charts showing various combinations of accuracy and precision. \textbf{b,} Schematic of the QM+QC framework. The quantum sensor is initialized in a probe state \ket{\psi_0}, evolves under noisy sensing superoperator $\tilde{\mathcal{U}}_\phi$ to yield a mixed state \(\tilde{\rho}_{t}\), which is then transferred to a quantum computer for optimization. The optimized state becomes closer to the ideal target, thus improving both accuracy and precision. \textbf{c,} Accuracy enhancement ($\Delta F$) achieved by the QM+QC method compared with standard QM. Accuracy is characterized by the fidelity between the measured and ideal states; the gray dashed line marks $\Delta F = 0$. Error probability is $1-P_0$ and noise strength is $1-\mathrm{Tr}[\rho_t \tilde{N}(\rho_t)]$. \textbf{d,} Comparison of QFI for a 6-qubit GHZ state with and without optimization, showing a 13 dB improvement in QFI at $P_{0} = 0.5$.}
		\label{fig1}
	\end{figure*}
	
\section*{Results}
\subsection*{Framework of noisy quantum metrology}
The primary goal of QM is to estimate unknown physical parameters with a precision surpassing classical limits. As illustrate in Fig.~\ref{fig1}b, a typical QM protocol begins by initializing the system in a probe state $\rho_0 = |\psi_0\rangle\langle \psi_0|$, where $|\psi_0\rangle$ may exhibit entanglement to enhance measurement sensitivity. In magnetic field sensing, for instance, the objective is to estimate an unknown field strength by determining the corresponding frequency $\omega$. 
As the probe evolves over time $t$, a phase $\phi = \omega t$ is imprinted on the state, yielding $\rho_t = |\psi_t\rangle\langle \psi_t|$ with $|\psi_t\rangle = U_\phi |\psi_0\rangle = e^{-i\phi} |\psi_0\rangle$. Measurements are then performed on $\rho_t$ to infer $\phi$, and thereby $\omega$. When probes are used independently, the estimation precision is bounded by the standard quantum limit. By contrast, entangled states, such as the Greenberger–Horne–Zeilinger (GHZ) state, can surpass the standard quantum limit and approach the ultimate Heisenberg limit. In QM, two key performance indicators are accuracy, which reflects the closeness to the true value, and precision, which quantifies the reproducibility of measurements.

In realistic settings, however, quantum sensors are inevitably affected by noise. During the sensing process, interactions with the environment introduce deviations in the measurement outcomes. This noisy evolution can be modeled by a superoperator $\tilde{\mathcal{U}}_{\phi} = \Lambda \circ U_{\phi}$, where $\Lambda$ denotes the noise channel. Throughout this work, we use the tilde notation $\tilde{M}$ to indicate that an operator $M$ is noise-corrupted.

The simplest scenario considers a two-stage evolution process (see \textit{Methods}), leading to the final state:
\begin{equation}
    \tilde{\rho}_t = \tilde{\mathcal{U}}_\phi(\rho_0) = \Lambda(\rho_t) = P_0 \rho_t + (1 - P_0)\tilde{N} \rho_t \tilde{N}^\dagger,
    \label{eq1}
\end{equation}
where $P_0$ is the probability of no error, and $\tilde{N}$ is a unitary noise operator. Such environmental noise degrades both the accuracy and precision of metrological tasks. To mitigate these effects, the central challenge becomes extracting as much useful information about $\rho_t$ as possible from the noise-corrupted state $\tilde{\rho}_t$.

\subsection*{Optimization using quantum computing}
To achieve noise-resilient QM in realistic settings, we process the noise-corrupted state $\tilde{\rho}_{t}$ on a quantum computer to extract and optimize its informative content. This transfer stage can be realized using standard quantum techniques such as quantum state transfer \cite{matsukevich2004quantum,chaneliere2005storage,yuan2008experimental,xiang2024enhanced} or quantum teleportation \cite{olmschenk2009quantum,ren2017ground}, which avoid the classical data-loading bottleneck since $\tilde{\rho}_{t}$ remains a quantum state. We assume no additional errors are introduced during this step.
Once $\tilde{\rho}_{t}$ is encoded on a quantum processor, we apply quantum machine learning techniques to extract useful information. In particular, we employ quantum principal component analysis (qPCA) \cite{lloyd2014quantum}, which leverages quantum computational advantages to efficiently extract the dominant components of noise-corrupted quantum states.

Experimentally, qPCA can be implemented via multiple copies of the input state~\cite{lloyd2014quantum}, repeated state evolutions~\cite{lin2024hardware}, or variational quantum algorithms~\cite{xin2021experimental}. Platforms that have realized qPCA include superconducting circuits \cite{kjaergaard2022demonstration}, NV centers in diamond~\cite{li2021resonant}, and nuclear magnetic resonance~\cite{lin2024hardware,xin2021experimental}. Detailed implementation procedures are provided in the \textit{Methods}. In this work, both the NV-center experiment and simulations on distributed superconducting processors adopt the variational approach.
We denote the resulting state after qPCA as $\rho_{\text{NR}}$, where the subscript means this state is more noise-resilient. To quantify the performance enhancement, we compare the accuracy and precision of $\rho_{\text{NR}}$ with those of the original noise-corrupted state $\tilde{\rho}_{t}$.

For accuracy, we compute the fidelity with respect to the ideal target state $|\psi_t\rangle$. Specifically, the fidelities are given by $\tilde{F} = \langle \psi_t | \tilde{\rho}_{t} | \psi_t \rangle$ and $F = \langle \psi_t | \rho_{\text{NR}} | \psi_t \rangle$, before and after optimization, respectively. In Fig.~\ref{fig1}c, we show the fidelity enhancement $\Delta F = F - \tilde{F}$ as a function of the error probability $1-P_0$ and the noise strength, quantified by $1-\mathrm{Tr}[\rho_t \tilde{N}(\rho_t)]$, which captures the deviation induced by the noise operator. The results show that when $P_0 > 0.5$, we always observe $\Delta F > 0$, indicating QM+QC consistently improves accuracy regardless of noise strength. This regime is typical in realistic metrology tasks. Even in the regime $P_0 < 0.5$, QM+QC can still improve accuracy as long as the noise operator is not too strong, i.e., when $\mathrm{Tr}[\rho_t \tilde{N}(\rho_t)] > 0.5$. In both scenarios, QM+QC yields a state $\rho_{\text{NR}}$ that achieves superior accuracy compared with the noise-corrupted state $\tilde{\rho}_{t}$.

Beyond accuracy, QM+QC also enhances measurement precision, which is quantified by the QFI, which sets the ultimate bound on parameter estimation via the quantum Cramér--Rao inequality~\cite{degen2017quantum}. We consider a 6-qubit GHZ probe state, which ideally attain the Heisenberg limit, and evaluate the QFI before and after QC optimization under varying noise levels. As shown in Fig.~\ref{fig1}d (see also \textit{Methods} and \textit{Supplementary Information}), QM+QC leads to a clear enhancement in QFI. These results confirm that our approach provides an effective pathway toward noise-resilient metrology across a broad range of practical scenarios.	

\subsection{Sensing magnetic field using NV centers}
	
	\begin{figure*}
		\centering
		\includegraphics[width=0.95\linewidth]{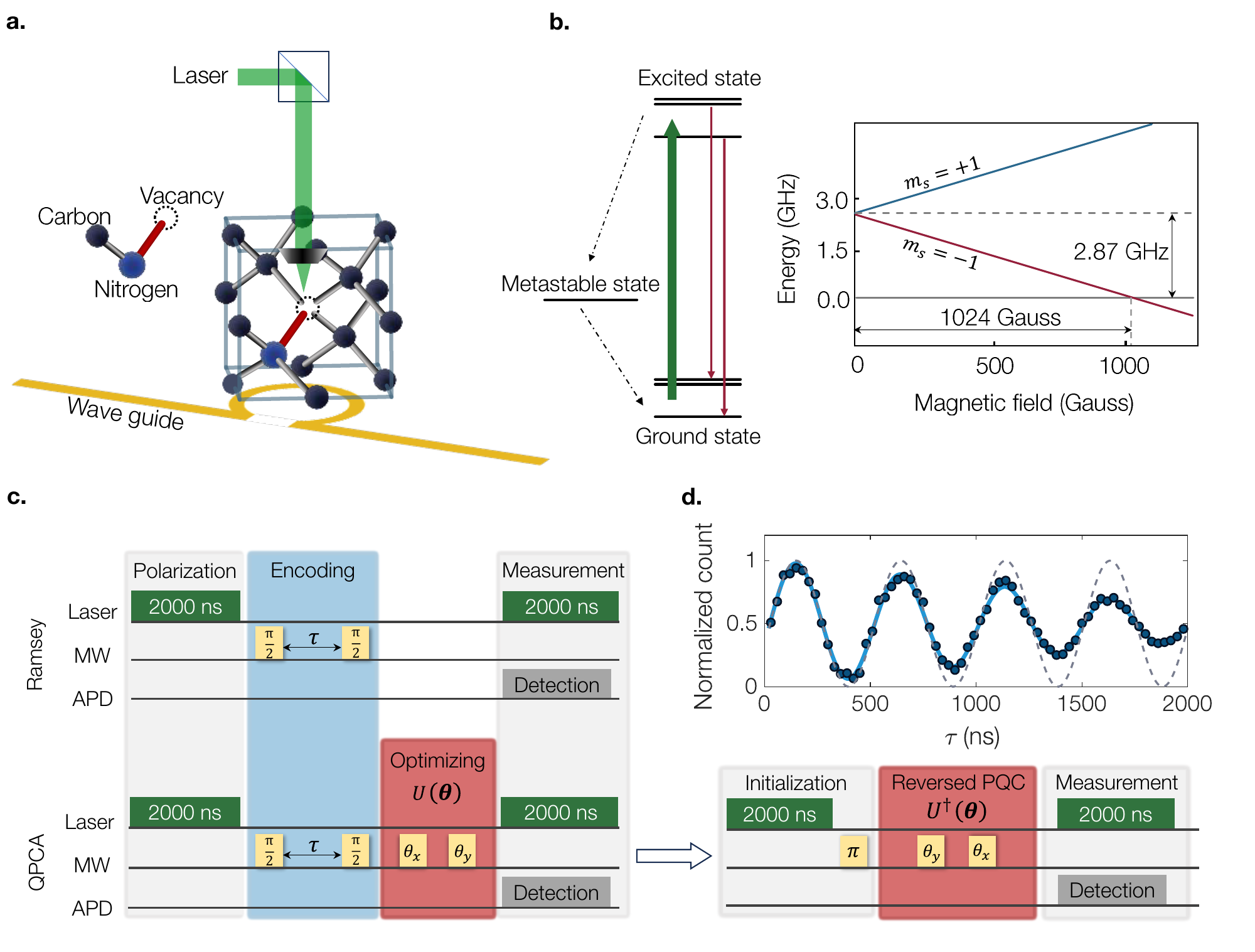}
		\caption{\textbf{Experimental setup of the NV-center platform and implementation of the QM+QC protocol.} \textbf{a,} Experimental setup and schematic structure of the NV center in diamond. Black spheres represent carbon atoms, the blue sphere denotes the nitrogen atom, and the dashed circle marks the vacancy. The yellow curve indicates the microwave waveguide. \textbf{b,} Energy-level diagram of the NV center. Green laser excitation initializes the electron spin, and red fluorescence from the excited- to ground-state transition enables spin-state readout. The ground triplet state \(m_s= 0,\pm 1\) exhibits a 2.87 GHz zero-field splitting, and an applied magnetic field lifts the \(m_s= \pm 1\) degeneracy. \textbf{c,} Pulse sequence of the Ramsey and qPCA experiments. A 2000 ns laser pulse polarizes the spin into \ket{0} state; two microwave $\pi/2$ pulses separated by $\tau$ encode magnetic-field information. For the qPCA sequence, additional parameterized microwave pulses $U(\bm{\theta})$ are applied and optimized; after convergence, the inverse operation $U^{\dagger}(\bm{\theta})$ retrieves the optimized sensing result. \textbf{d,} Ramsey fringes measured under noise. Blue dots represent experimental data fitted by the solid blue curve, while the gray dashed line shows the ideal Ramsey signal.}
		\label{fig2}
	\end{figure*}

Next, we validate our approach using NV centers in diamond. The NV center is a well-established solid-state quantum system with broad applications across quantum technologies \cite{shi2010room,pompili2021realization,randall2021many,shi2015single,hsieh2019imaging}. It is a promising candidate for advancing QM from laboratory-scale research to practical, industrial applications~\cite{degen2017quantum}, and the high-precision control capabilities also give this system the ability to perform QC tasks. It is a point defect in diamond, as shown in Fig.~\ref{fig2}a, where its energy-level structure is shown in Fig.~\ref{fig2}b. Laser excitation can polarize the spin and by collecting the fluorescence emitted from the excited state to the ground state one can read out the spin state. The ground-state manifold is a spin triplet, the \(m_s=\pm 1\) sublevels are degenerate and split from the \(m_s=0\) sublevel by \(2.87\)~GHz. Because the spin levels are sensitive to magnetic fields, an applied field readily lifts the degeneracy and can thus be detected. The experimental setup of our system is described in \textit{Methods}.

In our experiment, we first employ the single electron spin as the metrological probe to detect a magnetic field in the presence of noise. We set the target magnetic field \( B_{0} = 0.25 \) Gauss, and measure it using the Ramsey method. During the free evolution \( \tau \), the probe accumulates the phase \( \phi =  \gamma_e B \tau \), where \( \gamma_e \) is the gyromagnetic ratio of the electron spin. A subsequent \( \pi/2 \) pulse is then applied to map this phase onto the population of \( |0\rangle \), enabling optical readout.

The noise is introduced using the Gaussian noise model \( \mathcal{N}(0, \sigma^{2}) \), where \( \mathcal{N} \) denotes a Gaussian distribution and \( \sigma \) is the noise strength in percentage. This modeling is consistent with the noise model in Eq. (\ref{eq1}), which induces an additional decay in the Ramsey signal (see Fig.~\ref{fig2}d), thereby shifting the correlation between the measured population and the phase \( \phi \). This leads to sizable errors between the sensing estimates and the true $B_0$. Consequently, this degradation reduces the accuracy of magnetic field estimation.

\begin{figure*}
	\centering
	\includegraphics[width=0.8\linewidth]{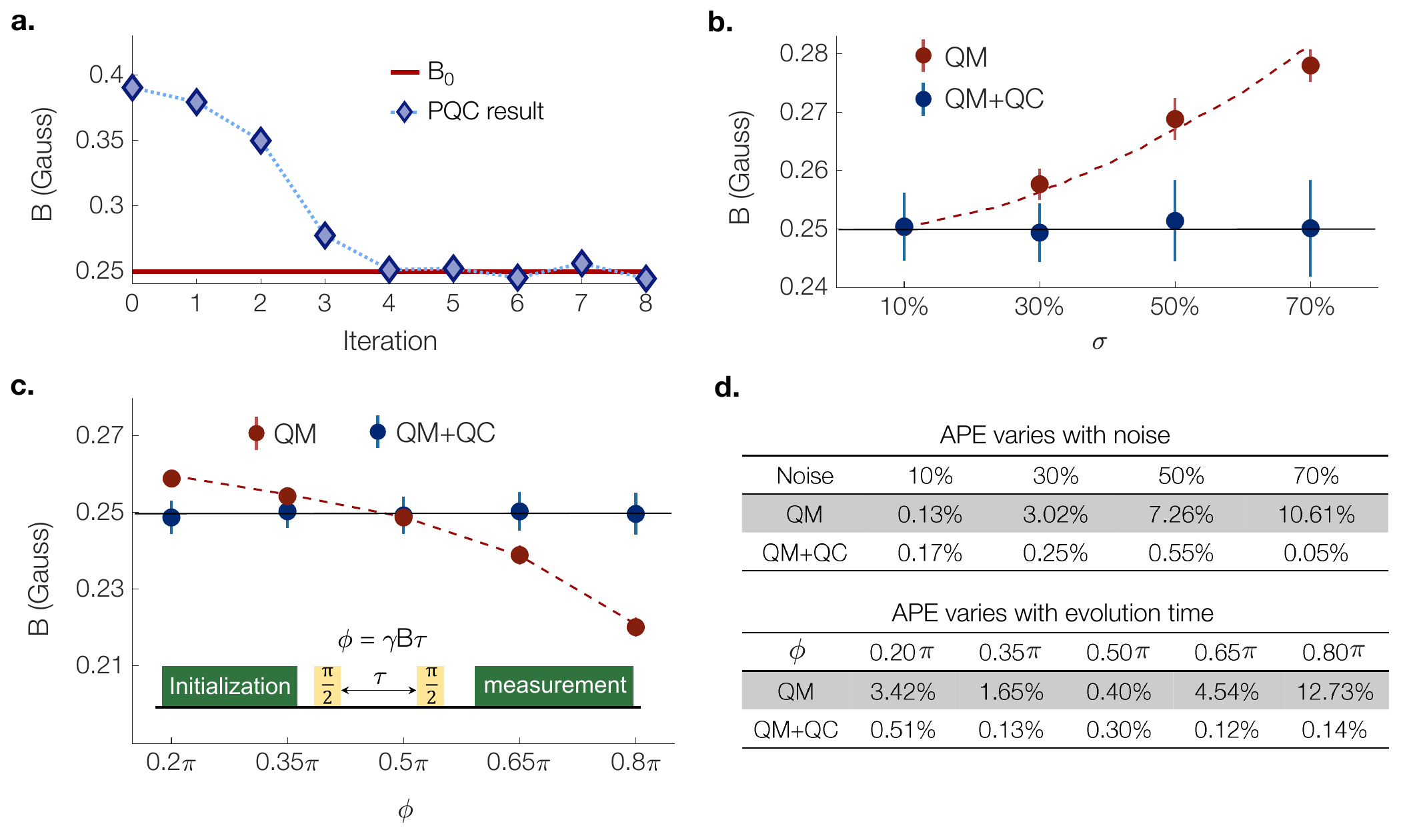}
	\caption{\textbf{Experimental results of the QM+QC protocol on NV centers.} \textbf{a,} Convergence of the measured magnetic field during iterative optimization. Blue diamonds show the optimized results approaching the target field $B_{0} = 0.25$~Gauss (red line). \textbf{b,} Magnetic-field sensing under different noise strengths ($\sigma$). Red and blue dots denote results from conventional Ramsey metrology (QM) and the QM+QC protocol, respectively. The red dashed line represents numerical simulation. \textbf{c,} Sensing results at various phases $\phi$ under $\sigma = 30\%$. QM+QC yields robust accuracy independent of evolution time, whereas the Ramsey method performs well only near the optimal phase $\phi = \pi/2$. \textbf{d,} APE for QM and QM+QC at various $\sigma$ and $\phi$, showing over 200-fold improvement in accuracy with the QM+QC method.}
	\label{fig3}
\end{figure*}

Next, we verify the performance of the QM+QC approach. Following the Ramsey sequence, the additional readout step is not immediately performed. Alternatively, we employ a parameterized quantum circuit (PQC)-based qPCA optimization sequence~\cite{xin2021experimental} to extracts the dominant component from the noise-corrupted state. The experimental pulse sequence is depicted in Fig.~\ref{fig2}c. The PQC consists of two rotational gates about the $x$- and $y$-axes, with corresponding tunable parameters $\theta_x$ and $\theta_y$. In experiment, we set the initial values of $\theta_x$ and $\theta_y$ to $\pi/2$ and $\pi/3$, and update them iteratively by measuring the gradient of the loss function; see Methods. With the final $\theta_x$ and $\theta_y$, we obtain the magnetic field in the following way: applying the reversed PQC sequence on $\ket{1}$, measuring the $\ket{0}$ population of the output, and mapping the result to the magnetic field.

In experiment, we monitor the measured magnetic field after each iteration. The results after eight iterations are shown in Fig.~\ref{fig3}a, where we see that the measured magnetic field eventually converges toward the target value $B_{0}=0.25$ Gauss in just a few iterations. 

We next assess its accuracy. To this end, we compare the measurement results obtained with and without QM+QC. As shown in Fig.~\ref{fig3}b, the results obtained via the traditional Ramsey method increasingly deviate from the true $B_{0}$ as noise intensity grows. In comparison, the estimates from the QM+QC approach display significantly reduced deviations. In our experiment, at the highest noise strength (\(\sigma = 70\%\)), the absolute percentage errors (APE, defined as $\left|\frac{B-B_{0}}{(B+B_{0})/2}\right|$) of the Ramsey and QM+QC results are \(10.16\%\) and \(0.05\%\) (see Fig.~\ref{fig3}d), respectively, corresponding to an accuracy improvement by a factor of more than \(200\) times.

We further show that QM+QC approach enables better performance at different measurement times \(\tau\). In the traditional Ramsey method, one typically aims to perform sensing tasks at the optimal measurement time, which gives \(\phi = \pi/2\), for the best sensitivity. However, in real-world sensing applications, the true value of the magnetic field to be measured is generally unknown, making it difficult to exploit this optimal measurement time. We show that QM+QC is independent of the measurement time. Even when far from the optimal time, we still observe remarkable accuracy improvements, as shown in Fig.~\ref{fig3}c. For example, at a noise strength of \(\sigma = 30\%\), the Ramsey method performs well only at the optimal time, whereas the QM+QC approach consistently converges to the true magnetic field across different measurement times. The inset compares the APE of the two methods under different phase accumulations. As shown in Fig.~\ref{fig3}d, at \(\phi = \pi/2\), the APE using Ramsey is small ($0.40\%$), but rises to \(12.73\%\) while away from the optimal point. By contrast, the APEs of QM+QC are always below \(0.60\%\). This result highlights the stability against measurement times using our method.

\subsection{Sensing magnetic field using superconducting circuits}
The NV center system is a promising quantum sensor using the electron spin as the probe. However, it is challenging to produce entangled states and hence reach the quantum advantage at Heisenberg limit. Alternatively, superconducting circuits have shown great capabilities in generating multi-qubit entangled states and become a promising candidate for QM. Here we show the applicability of our method in dealing with noise-corrupted entangled states on the superconducting platform.

    \begin{figure*}
		\centering
		\includegraphics[width=0.95\linewidth]{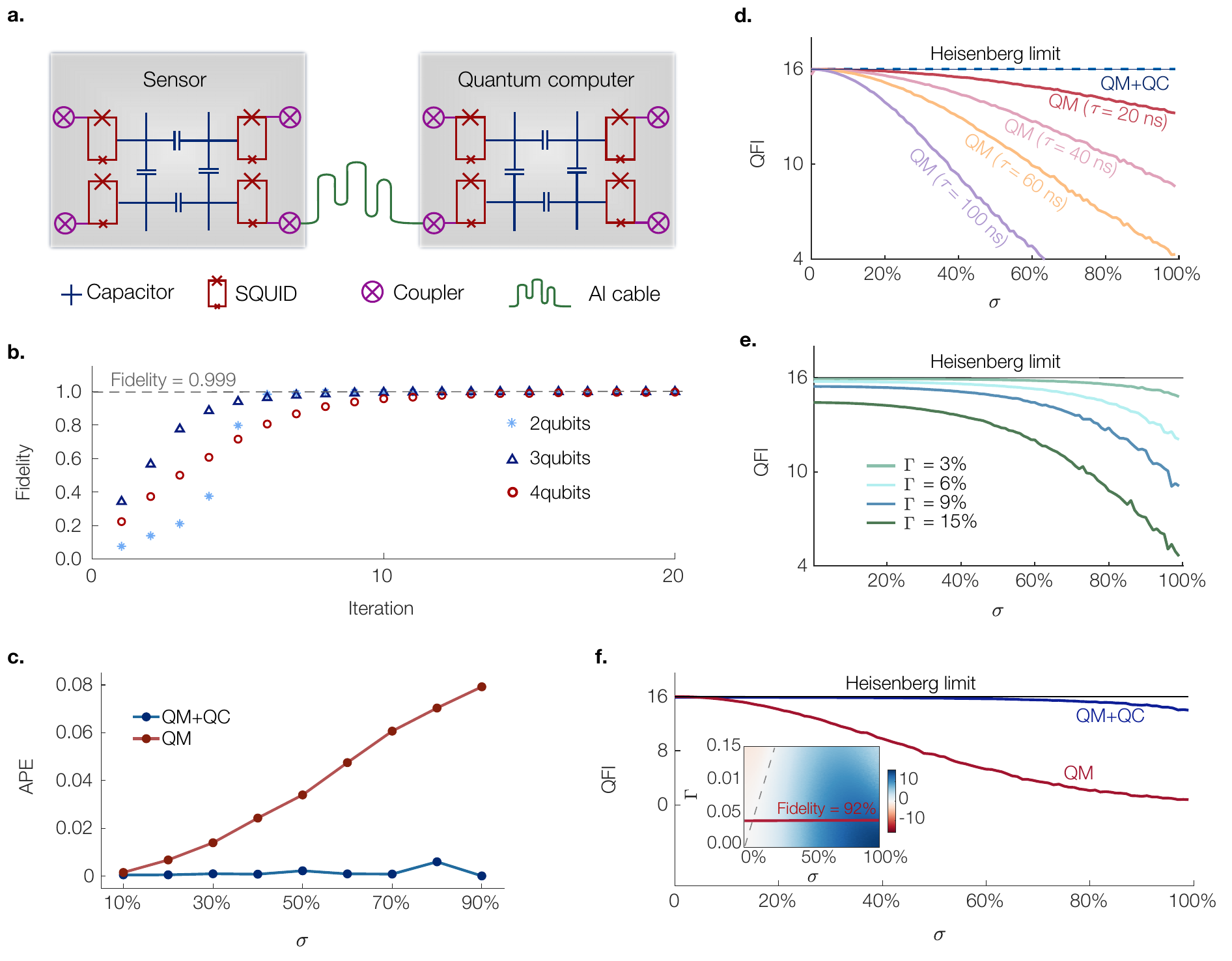}
		\caption{\textbf{QM+QC approach using distributed superconducting chips.} \textbf{a,} Modular superconducting schematic comprising two four-qubit chips interconnected by a low-loss aluminum coaxial cable. One module acts as the quantum sensor, the other as the quantum computer. \textbf{b,} Evolution of state fidelity during PQC optimization for 2-, 3-, and 4-qubit GHZ probes. \textbf{c,} APE versus noise strength. QM+QC maintains APE $<0.6\%$, while conventional QM shows significant deviation at high noise levels. \textbf{d,} QFI as a function of noise strength for evolution times $\tau = 20 \sim 100$ ns. The QM+QC results are shown for $\tau = 100$ ns (thin solid line), lying almost indistinguishably close to the Heisenberg limit (dashed line). \textbf{e,} QFI of the QM+QC method when the sensing results are transmitted through a noisy quantum channel modeled by amplitude-damping noise with loss rates $\Gamma = 3\% \sim 15\%$. \textbf{f,} QFI comparison between QM and QM+QC when the transmission fidelity is $92\%$ ($\Gamma = 4.15\%$), The inset shows the QFI enhancement $\mathcal{F}_{\text{QM+QC}} - \mathcal{F}_{\text{QM}}$ as a function of noise strength and channel loss, highlighting consistent improvement across realistic operating conditions.}
		\label{fig4}
	\end{figure*}

We consider the distributed architecture of superconducting chips, so that one module can be employed as the sensor and the other as the quantum computer for optimization. All numerical simulations are based on the real setting of a recently reported distributed superconducting chips in Ref.~\cite{niu2023low}, where five four-qubit quantum modules were successfully interconnected using pure aluminum coaxial cables with high-fidelity of generating the GHZ entangled states. This architecture also enable the coherent transmission of GHZ states across modules, which is most suitable for implementing our QM+QC approach.


We consider two interconnected modules, each consisting of four qubits (see Fig.~\ref{fig4}a). One module is used to prepare a four-qubit GHZ state that serves as the probe for magnetic field sensing, while the other functions as a quantum processor for optimization. The accumulated phase in a $n$-qubit GHZ probe is $\phi = n g_1 \Phi_{B_0} \tau$, where $\tau$ is the evolution time, $g_1$ denotes the linear response coefficient (first-order slope) and $\Phi_{B_0}$ represents the effective flux; see Methods.

We set the target magnetic field to \(B_0 = 0.16\) Guass, and take \(g_1A_{eff} = 42\) GHz/T, where $A_{eff}$ is the effective mutual inductance and $\Phi_{B} = A_{eff}B$. After the probe completes the sensing stage, we do not perform readout; instead, we transmit the entangled quantum state through a coaxial cable to the second module that implements optimization. As this computing module does not directly interface with the external magnetic field, it can operate with a lower error rate. The QM+QC approach is realized via the PQC of 16 tunable parameters. The parameters are optimized over multiple iterations until the loss function converges. After each iteration, the learned parameters define a unitary multi-qubit operation; by applying the inverse of this operation on the input state, we can read out the phase and subsequently obtain the magnetic field for that iteration. As shown in Fig.~\ref{fig4}b, for \(n=2,3,4\), the fidelities to the ideal (noise-free) target state reach \(99.9\%\) after 9, 12, and 27 iterations, respectively. 

We next access accuracy using the example a four-qubit GHZ probe. With the sensing time fixed at \(\tau=100~\mathrm{ns}\), Fig.~\ref{fig4}c shows their APE as a function of noise strength for both approaches. For QM+QC approach, the APE remains below \(0.60\%\) and is typically around \(0.10\%\). By contrast, the traditional method deviates increasingly from the true value as the noise grows, with the APE reaching as high as \(7.92\%\).

For precision, we characterized it using QFI. Specifically, we study the behavior of the QFI under different sensing durations and Gaussian noise levels. The corresponding results are shown in Fig.~\ref{fig4}d. It can be seen that, at larger noise strengths, the QFI decays rapidly with increasing measurement time. In contrast, the QFI in our approach lies very close to the Heisenberg limit, which shows only a negligible dependence on the noise level.

Since fidelity loss is inevitable during quantum state transfer, we further figure out if the QM+QC approach remain advantageous in the presence of transmission-induced error. To address this, we assume a noise channel that simulates transmission loss, described by the map $\rho \rightarrow E_0 \rho E_0^\dagger + E_1 \rho E_1^\dagger$, where
\begin{equation}
E_0 = \begin{bmatrix}
1 & 0 \\
0 & \sqrt{1 - \Gamma}
\end{bmatrix}, \quad
E_1 = \begin{bmatrix}
0 & \sqrt{\Gamma} \\
0 & 0
\end{bmatrix},
\end{equation}
and $\Gamma$ is the channel loss rate. This channel corresponds to the amplitude damping channel, commonly used to model information loss during transmission through coaxial cables.

Fig.~\ref{fig4}e presents the QM+QC results with the transmission noise channel, with \(\tau = 20\)~ns. We show the cases \(\Gamma = 3\%, 6\%, 9\%,\) and \(15\%\). When the transmission loss is large, the noise-resilience of QM+QC indeed declines rapidly and can approach the standard quantum limit. Even though, when compared with the traditional method QM+QC still demonstrates significant improvement across most parameter regions, as shown in sub-figure inside Fig.~\ref{fig4}f. In extreme cases ($\sigma = 10\%$ and $\Gamma = 0$), we have  the QFI at $0.754$ without optimization, which rises to $15.995$ after optimization, indicating an improved precision by 13.27 dB. In particular, we highlight in red the data point corresponding to a transmission fidelity of \(92\%\), as reported in Ref.~\cite{niu2023low}, which shows a clearly visible enhancement, and show the QFI of two groups in Fig.~\ref{fig4}f. It means that, under the experimental error level, the improvement in QFI is consistently present using QM+QC for sensing tasks.

\section{Conclusion}
In conclusion, we demonstrate a noise-resilient QM+QC protocol for sensing tasks. We employ qPCA to mitigate decoherence errors induced by environmental noise. Theoretical analysis confirms that the optimization significantly improves both fidelity and QFI, thus enhancing the accuracy and precision of for a sensing task. Experimentally validated using the NV-center platform and and numerically simulated using the distributed superconducting processors, our results show that QM+QC effectively enhances the accuracy and precision when sensing a magnetic field. By utilizing quantum machine learning to directly process quantum datasets, our study highlights a promising pathway toward integrating quantum computation with sensing tasks, ultimately paving the way for broader applications of quantum computers.


\section{Methods}

\subsection{Accuracy and precision}
In metrology, accuracy and precision are two distinct yet complementary indicators of measurement performance. Accuracy characterizes how close the measured value is to the true or reference value, while precision quantifies the repeatability of measurements under identical conditions. In the context of quantum metrology, both metrics are essential for evaluating the performance of a sensing protocol.

Experimentally, accuracy is often affected by systematic errors and is therefore assessed through the deviation between the measured quantity and its true value. 
Here, we define accuracy using the absolute percentage error (APE),
\begin{equation}
    \mathrm{APE} = 
    \left|\frac{A - M}{(A + M)/2}\right|,
\end{equation}
where $A$ denotes the actual (true) value and $M$ the measured or predicted value. A smaller APE corresponds to higher measurement accuracy.

Precision, on the other hand, describes the statistical reliability of repeated measurements. In classical measurement theory, this concept corresponds to test–retest reliability and is fundamentally limited by random fluctuations in the measurement process. Even in the absence of systematic bias, the variance of the estimated parameter cannot be reduced arbitrarily. The theoretical lower bound of such uncertainty is given by the Cramér–Rao bound. 

In the quantum regime, measurements are described by positive operator-valued measures, and precision is fundamentally constrained by the quantum Cramér–Rao bound,
\begin{equation}
    \mathrm{Var}(\theta) \geq \frac{1}{N\,\mathcal{F}(\theta)},
\end{equation}
where $N$ is the number of repeated measurements and $\mathcal{F}(\theta)$ denotes the QFI. The QFI depends solely on the geometric structure of the quantum state family $\{\rho(\theta)\}$ and quantifies the ultimate attainable precision, independent of the specific measurement strategy. An enhanced $\mathcal{F}$ therefore directly reflects an improved measurement precision, which serves as one of the key performance indicators for our QM+QC protocol.
	
\subsection{Noisy metrology model}
Consider a two-stage evolution of metrology. The initial state $|\psi_0\rangle$ is evolved into $|\psi_{t}\rangle$ by a sensing process
characterized by operator $\hat{O}$, and then a quantum noise channel with its operator $\hat{N}$ act on state $|\psi\rangle$, and the final state thus becomes
\begin{equation}
\tilde{\rho_t}=P_{0}|\psi_{t}\rangle\langle\psi_{t}|+(1-P_{0})\tilde{N}|\psi_{t}\rangle\langle\psi_{t}|\tilde{N}^\dagger 
\end{equation} 
We expand the final state $\tilde{N}|\psi_t\rangle$ by its original state $|\psi_t\rangle$ and its perpendicular state $|\psi_t^{\perp}\rangle$ as
\begin{equation}
\tilde{N}|\psi_{t}\rangle=\alpha|\psi_{t}\rangle+\beta|\psi_{t}^\perp\rangle,
\end{equation}
with $\langle\psi^{\perp}|\psi\rangle = \langle\psi|\psi^{\perp}\rangle = 0$, then the final state can be expressed in the basis of $|\psi_t\rangle,|\psi_t^{\perp}\rangle$ and simultaneously be diagonalized:
\begin{equation}
	\begin{aligned}
		\tilde{\rho_t} & =\left[\begin{array}{cc}
			P_{0}+|\alpha|^{2}(1-P_{0}) & \alpha \beta^{*}(1-P_{0}) \\
			\alpha^{*} \beta(1-P_{0}) & |\beta|^{2}(1-P_{0})
		\end{array}\right] \\
			&=K\left(
		\begin{array}{cc}
				\lambda_{+} & 0 \\
				0 & \lambda_{-}
		\end{array}\right) K^{\dagger}\\
			&=K \Lambda_{f} K^{\dagger},
	\end{aligned}
	\label{eq3}
\end{equation}
where \(K=(\eta_{+}, \eta_{-})\), and \(\eta_{+}\) and \(\eta_{-}\) are the eigenvectors of the matrix $\tilde{\rho_t}$, with \(\lambda_{+}\) and \(\lambda_{-}\) being the corresponding eigenvalues. By retaining the larger principal component, we obtain \(\tilde{\rho}_{t+}\). The new fidelity is defined as
\begin{equation}
	\tilde{F}=\langle\psi_{t}|\tilde{\rho}_{t+}|\psi_{t}\rangle=|\langle+|\psi_{t}\rangle|^{2}=|\langle+|K|+\rangle|^{2}.\label{eq4}
\end{equation}
Here, \(|+\rangle=K^{\dagger}|\psi_{t}\rangle\) represents the state corresponding to the larger eigenvalue. It is worth noting that the \(\tilde{\rho}_{t+}\) here and the optimized quantum state \(\rho_{\mathrm{NR}}\) are equivalent in computation since the analysis involves direct extraction of the principal component of $\tilde{\rho_t}$.

For QFI calculation, we use the spectral decomposition of the density matrix, as the quantum state we aim to processing is mix state in some situation, where the density operator is expressed in terms of its eigenvalues and eigenvectors: $\rho(\xi) = \sum_{k} p_k(\xi)\ket{\psi_k(\xi)}\bra{\psi_k(\xi)}$, where $p_k$ are the eigenvalues and \ket{\psi_k} are the corresponding eigenvectors. Then the QFI can be evaluated:
\begin{align}
    \mathcal{F}[\rho(\xi)] = &\sum_{k: p_k>0} \frac{\left(\partial_\xi p_k\right)^2}{p_k}\\
    &+2 \sum_{k, l: p_k+p_l>0} \frac{\left(p_k-p_l\right)^2}{p_k+p_l}\left|\left\langle\psi_k \mid \partial_\xi \psi_l\right\rangle\right|^2,
\end{align}
 where $\xi$ is the parameter used by QFI to quantify the sensitivity of the $\rho$ to its. In our model, $\xi$ is the quantum phase $\phi$ obtained by encoding magnetic field information into the quantum state in the QM experiment.

\subsection{Quantum principal component analysis}
Principal component analysis (PCA) is a fundamental unsupervised learning algorithm that identifies the dominant components of a dataset by projecting high-dimensional data onto a lower-dimensional subspace while retaining maximal variance. In classical computation, the core of PCA lies in diagonalizing the covariance matrix
\begin{equation}
    C = \frac{1}{M}\tilde{X}\tilde{X}^{\top},
\end{equation}
where $\tilde{X}$ is the normalized data matrix constructed from $M$ samples. The eigenvalues and eigenvectors of $C$ are then obtained, and the leading eigenvectors are used to reconstruct the most informative subspace. This eigen-decomposition step dominates the computational cost of classical PCA, scaling as $O(N^2)$ for an $N$-dimensional feature space.

QPCA provides an exponential improvement in efficiency, with a query complexity that scales as $O[(\log d)^2]$ for a quantum system of dimension $d$. Because the covariance matrix is Hermitian and positive semidefinite, it can be represented by a quantum density matrix $\rho$, allowing PCA to be performed directly on quantum data. The goal of qPCA is to extract the principal eigencomponents of $\rho$, that is, to identify the most significant subspace of a mixed quantum state—without classical reconstruction. This capability makes qPCA particularly suitable for our purpose of processing noise-corrupted quantum states obtained from quantum metrology.

Theoretically, qPCA diagonalizes $\rho$ using quantum phase estimation, where the unitary evolution $U = e^{-i\rho t}$ is implemented on multiple copies of the state. A practical realization is achieved through the controlled-swap scheme, which satisfies
\begin{equation}
    \mathrm{Tr}_P\!\left(e^{-iS\Delta t}\, \sigma \otimes \rho\, e^{iS\Delta t}\right)= e^{-i\rho \Delta t}\, \sigma\, e^{i\rho \Delta t} + O(\Delta t^2),
\end{equation}
where $S$ is the swap operator and $\mathrm{Tr}_P$ denotes the partial trace over the ancillary subsystem. Repeating this operation on $n$ copies of $\rho$ effectively implements $e^{-i\rho n\Delta t}\sigma e^{i\rho n\Delta t},$ from which the eigenvalues and eigenvectors of $\rho$ can be efficiently extracted~\cite{lloyd2014quantum}. This provides the theoretical foundation for our QM+QC protocol.

\subsection{Iterative optimization}
	
In our work, the qPCA is realized through a PQC technique, that variationally approximates the diagonalization of $\rho$. This variational qPCA requires only a few copies of the input state and is experimentally feasible on current quantum processors. By iteratively updating the circuit parameters, the PQC extracts the dominant eigencomponent of the noise-corrupted state $\tilde{\rho}_t$, producing a purified and more noise-resilient state $\rho_{\mathrm{NR}}$. This process effectively implements a quantum-native noise-mitigation protocol that preserves the quantum coherence essential for precision enhancement in our QM+QC framework.

The core idea of this implementation is to construct a parameterized unitary operator $U(\boldsymbol{\theta})$ that approximately diagonalizes the density matrix through repeated optimization. Starting from a random or heuristic initial parameter set~$(\boldsymbol{\theta}_{0})$, the parameters are iteratively refined according to a loss function~$L(\boldsymbol{\theta})$, which quantifies the deviation of the transformed density matrix $U(\boldsymbol{\theta})\tilde{\rho}_t U^{\dagger}(\boldsymbol{\theta})$ from a diagonal form. Minimizing this loss corresponds to extracting the leading principal component of $\tilde{\rho}_t$, thereby realizing the effect of qPCA without the need for explicit eigenvalue decomposition.

In each iteration, the gradient of the loss function with respect to the $j$-th parameter is estimated using the parameter-shift rule,
\begin{equation}
    g[\theta_j] = \frac{1}{2}\big[L(\boldsymbol{\theta}:\theta_j+\delta) - L(\boldsymbol{\theta}:\theta_j-\delta)\big],
\end{equation}
and the parameters are updated as
\begin{equation}
    \theta_j \leftarrow \theta_j - \epsilon_j\, g[\theta_j],
\end{equation}
where $\epsilon_j$ denotes the learning rate or step size for the $j$-th parameter. 
The optimization continues until the loss function converges to its minimum, indicating that the PQC has effectively captured the principal subspace of $\tilde{\rho}_t$.

After convergence, the optimized unitary $U(\boldsymbol{\theta}_{opt})$ can be viewed as an operator that purifies the input state. 
Applying its inverse $U^{\dagger}(\boldsymbol{\theta}_{opt})$ to the basis state $|1\rangle$ reconstructs the optimized sensing result, from which the physical quantity of interest (e.g., the magnetic field) can be inferred. 
This iterative learning process corresponds precisely to the optimization sequence implemented in our NV-center experiment and the simulations of distributed superconducting processors described in the main text.

\subsection{NV-center setup}
	
Our experimental implementation is based on a home-built confocal microscope designed for optical detection and coherent control of single NV centers in diamond. A continuous-wave 532~nm green laser provides optical initialization and readout of the electron spin. The laser intensity is modulated using an acousto–optical modulator, allowing sub–10~ns switching to generate laser pulses for spin polarization and fluorescence detection. Photoluminescence emitted from the NV center is collected through an oil-immersion objective lens with a numerical aperture of~1.45 and detected by a single-photon counting avalanche photodiode.
	
The diamond sample containing NV centers is placed beneath permanent magnets that generate a static magnetic field of approximately~500~Gauss. This field shifts the spin-resonance frequency of the NV electronic ground state to around~1.4~GHz. Optically detected magnetic resonance is used to monitor the spin transition, exploiting the fluorescence contrast between the $m_s=0$ and $m_s=\pm1$ states.

Microwave signals for coherent spin control are generated by amplitude modulation of a 1.4~GHz carrier using an arbitrary waveform generator and an IQ mixer. The microwave signal is transmitted via coaxial cables and delivered to the NV centers through a coplanar waveguide. These pulses implement the Ramsey and PQC sequences described in the main text. The free-evolution period~$\tau$ determines the phase accumulation $\phi = \gamma_e B \tau$, where $\gamma_e$ is the electron gyromagnetic ratio and $B$ the local magnetic field to be estimated.
	
Magnetic-field noise is introduced artificially following a Gaussian distribution $\mathcal{N}(0,\sigma^2)$, where $\sigma$ represents the noise amplitude in percentage. Because dynamically varying the external magnetic field for each pulse sequence is experimentally challenging, we emulate noise by generating~800 random magnetic-field values on a computer. Each field setting is measured~10,000~times, corresponding to approximately~$2\times10^6$ projective measurements in total. The resulting ensemble of noisy quantum states~$\tilde{\rho}_t$ serves as the input for the PQC-based qPCA optimization, enabling direct experimental demonstration of the QM+QC noise-resilient protocol.

\subsection{Transmon qubit Hamiltonian}

We take the lowest two energy levels as \ket{0} and \ket{1}, and write the flux as:
\begin{equation}
\Phi = \Phi_{dc}+\Phi_{B}+\Phi_{ac}\cos{\Omega t},
\end{equation}
where $\Phi_{dc}$ is the flux generated by the applied bias current, \(\Phi_{B}\) is the flux induced by the magnetic field, and \(\Phi_{ac}\) is an additional modulation flux. The Hamiltonian can then be written as
\begin{equation}
H(t) = \frac{\hbar\omega_{01}(\Phi)}{2}\sigma_{z}+H_{\mathrm{ctrl}}(t),
\end{equation}
where \(H_{\mathrm{ctrl}}\) is the control Hamiltonian, \(\omega_{01}\) is the energy splitting between \ket{0} and \ket{1}, and
\begin{equation}
\hbar\omega_{01}(\Phi)\approx \sqrt{8E_{C}E_{J,eff}(\Phi)}-E_{C}.
\end{equation}
Expanding around a chosen \(\Phi_{dc}\), we obtain
\begin{equation}
\omega_{01}(\Phi) \approx \omega_0+g_1 \Phi(t)+\frac{1}{2} g_2  \Phi(t)^2+\cdots,
\end{equation}
where \(g_n=\partial^n \omega_{01} /\left.\partial \Phi^n\right|{\Phi_{\mathrm{dc}}}\) and \(\Phi(t)=\Phi_B+\Phi_{\mathrm{ac}} \cos \Omega t\). If we operate slightly off the sweet spot and set \(\Phi_{ac}=0\), then
\begin{equation}
H_{\mathrm{sig}} = \frac{\hbar}{2} g_1\Phi_{B}\sigma_{z}.
\end{equation}

\bibliography{Reference.bib}

\end{document}